\documentclass[10pt,conference]{IEEEtran}
\usepackage{algorithmicx}
\usepackage{algorithm}
\usepackage{subcaption}
\usepackage{diagbox}
\usepackage{geometry}
\usepackage{tabularx}
\usepackage{comment}
\usepackage{makecell}
\usepackage{fancyhdr} 
\usepackage{algpseudocode}
\usepackage{booktabs,multirow,tabularx}
\usepackage{bbm}
\usepackage{ulem}
\newtheorem{remark}{Remark}

\usepackage{setspace}
\makeatletter
\newcommand{\removelatexerror}{\let\@latex@error\@gobble}
\makeatother
\IEEEoverridecommandlockouts
\usepackage{cite}
\usepackage{amsmath,amssymb,amsfonts}
\usepackage{graphicx}
\usepackage{textcomp}
\usepackage{xcolor}
\def\BibTeX{{\rm B\kern-.05em{\sc i\kern-.025em b}\kern-.08em
		T\kern-.1667em\lower.7ex\hbox{E}\kern-.125emX}}

\begin{document}
\flushbottom
\algnewcommand\algorithmicswitch{\textbf{switch}}
\algnewcommand\algorithmiccase{\textbf{case}}
\algnewcommand\algorithmicassert{\texttt{assert}}
\algnewcommand\Assert[1]{\State \algorithmicassert(#1)}%
\algdef{SE}[SWITCH]{Switch}{EndSwitch}[1]{\algorithmicswitch\ #1\ \algorithmicdo}{\algorithmicend\ \algorithmicswitch}%
\algdef{SE}[CASE]{Case}{EndCase}[1]{\algorithmiccase\ #1}{\algorithmicend\ \algorithmiccase}%
\algtext*{EndSwitch}%
\algtext*{EndCase}%

\title{Multi-modal Data Driven Virtual Base Station Construction for Massive MIMO Beam Alignment
}

\author{
	\IEEEauthorblockN{
		Yijie Bian\IEEEauthorrefmark{1}\thanks{This work was supported in part by the Hong Kong Research Grant Council under the Area of Excellence (AoE) Scheme Grant No. AoE/E-601/22-R; and the National Natural Science Foundation of China (NSFC) under Grant 62301156.}, 
		Wei Guo\IEEEauthorrefmark{1}, 
		Jie Yang\IEEEauthorrefmark{2}\IEEEauthorrefmark{3}, 
		Shenghui Song\IEEEauthorrefmark{1}, \\
		Jun Zhang\IEEEauthorrefmark{1}, \IEEEmembership{Fellow, IEEE},
		Shi Jin\IEEEauthorrefmark{2}\IEEEauthorrefmark{4}, \IEEEmembership{Fellow, IEEE},
		and Khaled B. Letaief\IEEEauthorrefmark{1}, \IEEEmembership{Fellow, IEEE}\\
		} 
	\IEEEauthorblockA{\IEEEauthorrefmark{1}Department of Electronic and Computer Engineering, \\The Hong Kong University of Science and Technology, Hong Kong, China\\}

\IEEEauthorblockA{\IEEEauthorrefmark{2}Key Laboratory of Measurement and Control of Complex Systems of Engineering, Ministry of Education, \\Southeast University, Nanjing 210096, China\\}

\IEEEauthorblockA{\IEEEauthorrefmark{3}Frontiers Science Center for Mobile Information Communication and Security, \\Southeast University, Nanjing 210096, China\\}
	
	\IEEEauthorblockA{\IEEEauthorrefmark{4}School of Information Science and Engineering, Southeast University, Nanjing 210096, China}\vspace{-0.8 cm}

}

\maketitle

\begin{abstract}
Massive multiple-input multiple-output (MIMO) is a key enabler for the high data rates required by the sixth-generation networks, yet its performance hinges on effective beam management with low training overhead. This paper proposes an interpretable framework to tackle beam alignment in mixed line-of-sight (LoS) and non-line-of-sight (NLoS) propagation environments. Our approach utilizes multi-modal data to construct virtual base stations (VBSs), which are geometrically defined as mirror images of the base station across reflecting surfaces reconstructed from 3D LiDAR points. These VBSs provide a sparse and spatial representation of the dominant features of the wireless environment. Based on the constructed VBSs, we develop a VBS-assisted beam alignment scheme comprising coarse channel reconstruction followed by partial beam training. Numerical results demonstrate that the proposed method achieves near-optimal performance in terms of spectral efficiency.
\end{abstract}


\section{Introduction}
\label{Introduction}


To meet the demand for higher data rates in the sixth-generation (6G) networks, exploration of spectrum at increasingly elevated frequencies becomes necessary. Higher frequencies, like millimeter-wave (mmWave), sub-terahertz (sub-THz), and THz bands, are poised to be cornerstones of 6G, which will deliver ultra-high bandwidth for massive data throughput, minimal latency, enhanced connectivity, and support for advanced applications for both outdoor and indoor scenarios \cite{roadmap_6g,understand_sub_THz_channel}.

\textcolor{black}{To compensate for the high path loss under high frequency, massive multiple-input-multiple-output (MIMO) antenna arrays are adopted. \textcolor{black}{Given the increasing channel dimension, channel estimation requires prohibitively high overhead. Beam training based on a predefined codebook stands out as a promising approach, as it does not require estimating the channel explicitly while still achieving comparable performance for beam management. However, beam training still incurs considerable training overhead when performing exhaustive training \cite{tutorial_bm}. \textcolor{black}{The channel characterized by the propagation of electromagnetic waves through reflection, scattering and diffraction, etc., between the base station (BS) and user equipment (UE) highly depends on the transceiver locations and the surrounding environment \cite{smart_factory_ba}}. Thus, a promising strategy is to incorporate environmental sensing data, such as light detection and ranging (LiDAR) \cite{lidar_beam_prediction}, radio detection and ranging \cite{radar_beam_prediction}, images from red-green-blue (RGB) cameras \cite{camera_beam_prediction}, and location data \cite{gps_beam_alignment}, for efficient beam management.}}

\textcolor{black}{Existing studies on multi-modal data assisted beam alignment mainly focus on line-of-sight (LoS) scenarios.} In \cite{lidar_gps_beam_tracking}, the study adopts both 3D LiDAR points and global positioning system (GPS) data to assist beam tracking in vehicle-to-vehicle LoS scenario. 
Furthermore, the mmWave received signal strength indicator, RGB image, and GPS data are fused to assist robust beam tracking in the vehicle-to-infrastructure LoS scenario \cite{multimodal_beam_tracking}. 


\textcolor{black}{However, LoS based sensors, such as cameras, can perform poorly in obtaining non-line-of-sight (NLoS) information. A new insight into the relationship between the environment and the wireless communication channel is proposed based on the channel knowledge map (CKM) in \cite{ckm_beam_alignment}. With the help of localization technologies, the UE's location can be used to directly determine the optimal beam pair by looking up pre-collected measurements \cite{ckm_beam_alignment}.} \textcolor{black}{However, CKM-based methods require a large volume of wireless measurements and have limited grid resolution. Moreover, they can contain redundant information since they do not take full advantage of propagation features of the radio signal. With the increase of frequency, the propagation paths are more sensitive to the dynamic environment, \textcolor{black}{which means that certain channels measured in CKM can have unwanted features that are difficult to eliminate.}}

\textcolor{black}{To overcome the limitations of existing methods, we propose \textcolor{black}{an explicit and interpretable }approach that leverages multi-modal data with global sensing features to construct a virtual base station (VBS) set for beam alignment, different from virtual anchor points in \cite{hybrid_slam} which are usually related to smooth walls for indoor simultaneous localization and mapping scenarios. \textcolor{black}{The VBSs are defined via the image principle: for each reflective surface within the LoS coverage of the physical BS, a corresponding VBS is created as its mirror image behind that surface. If such VBS is treated like the physical BS, higher order VBSs with more mirrored times can be constructed similarly.} Each VBS indicates that there could exist a reflection path, which can be used as a supplementary communication path when LoS does not exist. The contributions of the study are summarized as follows,	
\textcolor{black}{
\begin{itemize}
	\item We propose a multi-modal assisted VBS construction method to sparsely represent wireless environment features. Preprocessed 3D LiDAR points are meshed to model complex building facades. Original VBS locations are computed from the BS position and mesh parameters. Adjacent VBSs are then clustered to derive the final VBS set, and their virtual line-of-sight (VLoS) coverage areas are computed using geometric projection of reflective meshes.
	\item Using constructed VBSs and the UE's location, we coarsely reconstruct the channel to assign importance weights to beam pairs for efficient beam training. Coarse channel parameters, including the absolute value of channel gain, angle-of-arrival (AoA), angle-of-departure (AoD) and distance, can be estimated via empirical path loss models and geometric relations, without using the 3D LiDAR points, reducing computational overhead.
	\item A partial beam training method with flexible beam training numbers refines beam pairs for massive MIMO alignment with low online partial beam training overhead, evaluated on real-world Openstreetmap (OSM) and ray-tracing datasets.
\end{itemize}}}

\section{System Model}
\label{system_model}
\begin{figure}[h]
	\centering
	\includegraphics[width=0.6\linewidth]{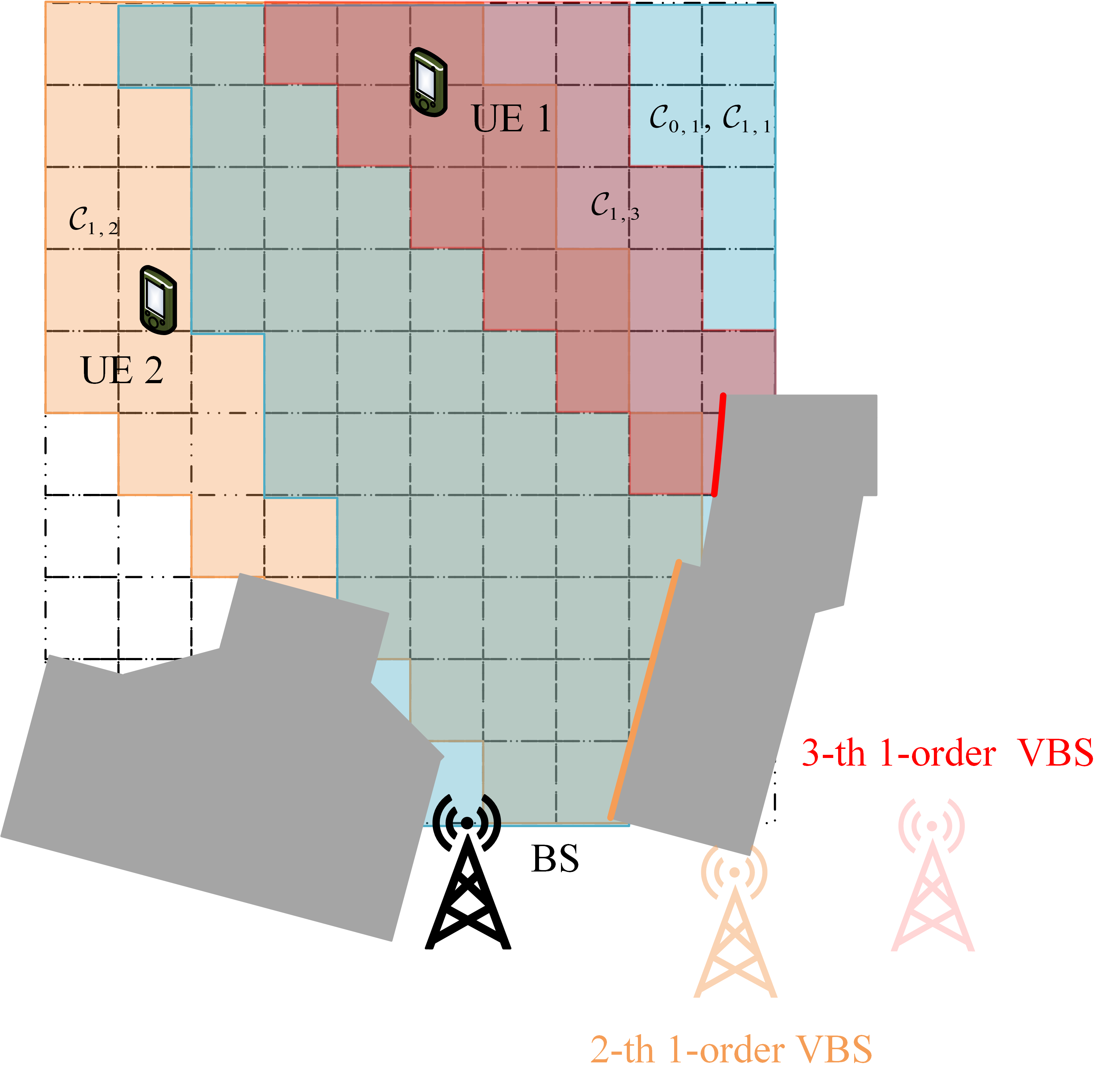}
	\caption{\textcolor{black}{Illustration of 1-order VBSs with corresponding coverage areas. With the ground reflection, there are 3 VBSs in the region. Apart from the shown 2 VBSs in red and orange, the VBSs with respect to the ground are below the physical BS and are not shown.}}
	\label{vbs_definition}
\end{figure}
\textcolor{black}{We consider a downlink massive MIMO communication system, where the BS is equipped with a uniform linear array (ULA) of $N_\mathrm{BS}$ transmit antennas, and the UE has $N_\mathrm{UE}$ receive antennas. The beamforming vectors are represented as $\mathbf{f} \in \mathbb{C}^{N_\mathrm{BS} \times 1}$ at the BS and $\mathbf{w} \in \mathbb{C}^{N_\mathrm{UE} \times 1}$ for the UE, respectively. \textcolor{black}{For ease of exposition, the BS and UE are assumed to have one RF chain}, the magnitude of each element in $\mathbf{f}$ and $\mathbf{w}$ is constant, satisfying $|\{\mathbf{f}\}_i| = {1}/{\sqrt{N_\mathrm{BS}}}$ and $|\{\mathbf{w}\}_j| = {1}/{\sqrt{N_\mathrm{UE}}}$ for $i = 1, \ldots, N_\mathrm{BS}$ and $j = 1, \ldots, N_\mathrm{UE}$. 
In terms of predefined codebook, due to the massive MIMO under high carrier frequency, a considerable portion of coverage area is in the near-field. Thus, we adopt the near-field polar-domain codebook matrix $ \mathbf{U}_\mathrm{BS}\in \mathbb{C}^{N_\mathrm{BS}\times M_\mathrm{BS}}, \mathbf{U}_\mathrm{UE}\in \mathbb{C}^{N_\mathrm{UE}\times M_\mathrm{UE}} $ for BS-side and UE-side, respectively. }

\textcolor{black}{The received signal at UE can be expressed as 
\begin{gather}
	y=\sqrt{P_\mathrm{T}}\mathbf{w}^\mathrm{H}\mathbf{H}\mathbf{f}s+\mathbf{w}^\mathrm{H}\mathbf{n},
\end{gather}
where $ \mathbf{H}\in\mathbb{C}^{N_\mathrm{UE}\times N_\mathrm{BS}} $ is the channel matrix,
$ P_\mathrm{T} $ is the BS transmit power, $ s $ is the signal with normalized power, and $ \mathbf{n}\in\mathbb{C}^{N_\mathrm{UE}\times 1} $ denotes the complex Gaussian noise in bandwidth $ W $ with noise power spectrum density $ N_\mathrm{0} $, i.e. $\mathbf{n}\sim \mathcal{CN}(0, N_\mathrm{0}W\mathbf{I}_{N_\mathrm{UE}})$. The spectral efficiency (SE) can be calculated as
\begin{gather}
	\mathrm{SE} = \log_2\left(1+\frac{P_\mathrm{T}|\mathbf{w}^\mathrm{H}\mathbf{H}\mathbf{f}|^2}{N_\mathrm{0}W} \right).
\end{gather}}

\textcolor{black}{Leveraging the geometric insights from out-of-band multi-modal data, we can identify a small set of promising propagation paths and their corresponding beam directions. This effectively prunes the searching space, reducing the problem to finding the best beam pair within a candidate subset $ \mathcal{S} $ with $ |\mathcal{S}| = S << M_\mathrm{BS}M_\mathrm{UE}$. Within the subset, the beam alignment objective is to find the optimal beam pair that can maximize the SE, which is equivalent to find the largest virtual channel gain $ |\mathbf{w}^\mathrm{H}\mathbf{H}\mathbf{f}| $ with given transmit power budget. Finally, the beam alignment problem can be formulated as
\begin{equation}
\begin{array}{ll}
\max\limits_{\mathbf{f},\,\mathbf{w}} & |\mathbf{w}^\mathrm{H}\mathbf{H}\mathbf{f}| \\
\text{s.t.} & \mathbf{f} = \{\mathbf{U}_\mathrm{BS}\}_{:,b_\mathrm{BS}} \\
            & \mathbf{w} = \{\mathbf{U}_\mathrm{UE}\}_{:,b_\mathrm{UE}} \\
            & (b_\mathrm{BS}, b_\mathrm{UE})\in \mathcal{S},
\end{array}
\label{beam_alignment_problem}
\end{equation}
where operation $ \{\cdot\}_{:,b} $ means selecting the $ b $-th column of the matrix, i.e., the $ b $-th codeword in the codebook.}

\section{VBS-assisted Beam Alignment}
\label{VBS_motivations}

\subsection{Motivations and Definition}

\textcolor{black}{With the higher carrier frequency, the cell coverage size shrinks, which means that sensing data with a global view of the cell, such as regional 3D LiDAR data, can be obtained either by measurement or reconstruction. At the same time, there exist a few dominant paths, such as the LoS path and a few reflection paths, while diffraction and scattering paths become weaker. In \cite{TERRA}, the authors proposed to use a ground reflection path to handle blockage of the LoS scenario. However, such ground reflection path may not be stable, which means such path may still have relatively high path loss since it could penetrate through part of the UE. This limitation highlights the need for alternative and more persistent NLoS paths, such as building facades, which can be used to provide stable reflection paths.
To geometrically model these dominant reflection paths, we introduce the concept of VBS, which will be detailed in the sequel.}

\textcolor{black}{In high frequency bands, such as mmWave and sub-THz, paths with multiple reflections usually exhibit low channel gains, which suggests that the contribution of such paths to the largest virtual channel gains, i.e., the objective function in problem (\ref{beam_alignment_problem}), is typically negligible. Therefore, the single-reflection path can be considered as the fundamental block for constructing the VBS. Analogous to a mirror effect, the reverse extension line of several single reflection paths that undergo one reflection with respect to a reflective surface can converge at a certain point. This point is defined as 1-order VBS, as the associated paths involve only one reflection. For higher order VBSs, the $ u $-order VBSs are defined as the mirror image of the $ (u-1) $-order VBSs corresponding to paths with $ (u-1) $ reflections. However, unlike indoor scenarios, where single reflections mostly come from smooth walls, the outdoor building facades have more complex features, making the direct identification of VBS considerably more challenging. To handle this difficulty, multi-modal data, i.e. regional 3D LiDAR points and BS's location are used to construct VBS in an explicit way, which is detailed in the sequel.}


	extcolor{black}{
Before proceeding, to formalize the VBS indexing, we define the following based on reflection conditions. The BS itself is defined as the unique 0-order VBS with LoS coverage area $\mathcal{C}_{0,1}$. The first 1-order VBSs are defined as the image of the BS reflected by the ground, indexed as $ (1,1) $ with VLoS coverage area $\mathcal{C}_{1,1}$. For consistency, we set $\mathcal{C}_{0,1} = \mathcal{C}_{1,1}$. As illustrated in the example scenario in Fig. \ref{vbs_definition}, UE 1 lies within the LoS coverage of the 0-order VBS (the direct LoS path) and VLoS coverage of three 1-order VBSs (the 1st, 2nd and 3rd), while the blocked UE 2 is served by only one 1-order VBS (the 2nd). More generally, the $ v $-th VBS of order $ u $, denoted as VBS $ (u, v) $, is characterized by two key properties: its geographical location $ \mathbf{o}_\mathrm{VBS}^{u,v} $ and its coverage area $ \mathcal{C}_{u,v} $. It is evident that as the order $ u $ increases, the number of potential VBSs may grow, but their individual coverage areas typically shrink. This occurs because the geometric conditions required for multiple consecutive reflections become significantly more restrictive, leading to smaller viable service regions.}
\begin{figure*}[t!]
	\centering
	\includegraphics[width=0.85\linewidth]{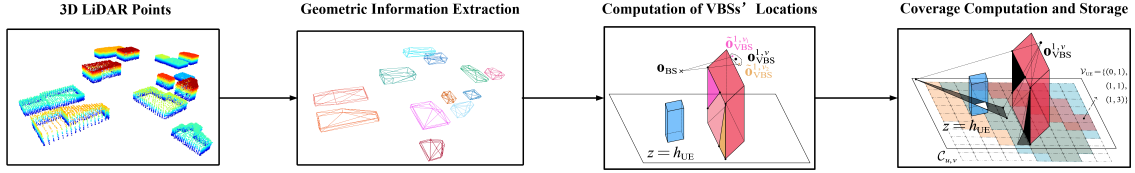}
	\caption{\textcolor{black}{Illustration of the VBS construction process.}}
	\label{VBS_construction}
\end{figure*}
\vspace{-0.0cm}
\subsection{VBS Construction and Storage}
\textcolor{black}{In this subsection, we introduce how to construct 1-order VBSs based on 3D LiDAR points and BS's location\footnote{Since $ u>1 $ order VBSs construction process is similar with the 1-st order VBSs, we omit the description of them due to space limitation.}. The VBSs construction process mainly includes three steps, which are geometric information extraction, computation of VBSs' locations, and coverage computation and storage, as shown in Fig. \ref{VBS_construction}.}

\subsubsection{Geometric Information Extraction}
\textcolor{black}{Recalling that the VBS definition is the mirror image of BS, the key problem here reduces to identify the building facades that function like the mirror. The complete facade of a building object may not be smooth enough to treat as a mirror, but the constituent meshes can be seen as independent mirrors. \textcolor{black}{The process of extracting these meshes begins with the acquisition of a regional 3D LiDAR point cloud. The LiDAR points can be obtained by measurement or reconstruction,} and are then preprocessed to remove noise, outliers, moving objects to preserve the static information of the environment.} Next, objects in the scene can be separated based on the LiDAR points by hierarchical density-based spatial clustering of applications with noise (HDBSCAN) algorithm \cite{HDBSCAN}. For the $m$-th object, the surface reconstruction \cite{alpha_shape} and surface simplification algorithm \cite{QEM} are applied to generate $ N_m $ triangular meshes set to represent the building object. \textcolor{black}{For the $ i $-th triangular mesh $ \mathcal{T}_i $, the outer normal vector $ \mathbf{r}_i\in \mathbb{R}^{3\times 1} $ can be computed based on the three edge vectors $ \mathbf{e}_i^k\in \mathbb{R}^{3\times 1}, k=1,2,3 $, where each edge vector can be computed by the adjacent vertex location $ \mathbf{v}_i^k\in \mathbb{R}^{3\times 1}, k=1,2,3 $. The centroid of the triangular mesh is denoted as $ \mathbf{v}_i^\mathrm{c}\in \mathbb{R}^{3\times 1} $, which can be computed as $ \mathbf{v}_i^\mathrm{c} = \frac{1}{3}\sum_{k=1}^3 \mathbf{v}_i^k $. These four points, together with the outer normal vector, can be used to geometrically characterize the triangular mesh.}

\subsubsection{Computation of VBSs' Locations}
\label{vbs_location_computation}
\textcolor{black}{The computation of VBSs' locations is not a simple reflection of the BS location $ \mathbf{o}_\mathrm{BS}\in \mathbb{R}^{3\times 1} $ across every triangular mesh in the region, which would generate a prohibitively large and redundant set of VBS. Instead, we introduce a systematic selection process to identify only the valid meshes that can produce meaningful, physically-realizable reflection paths. We posit that the $ i $-th triangular mesh is a valid reflector only if it satisfies both two fundamental geometric conditions, i.e.  orientation condition and unobstructed path condition.}
\begin{itemize}
	\item{\textbf{Orientation Condition:}} This condition ensures that the mesh is geometrically oriented towards the BS, a prerequisite for any reflection. Let $ \mathbf{p}_{i}^\mathrm{c} $ denote the vector pointing from BS to the centroid of mesh $ \mathcal{T}_i $, which is computed as $ \mathbf{p}_{i}^\mathrm{c}=\mathbf{v}_{i}^\mathrm{c}-\mathbf{o}_\mathrm{BS} $. The orientation condition can be expressed mathematically as
	\begin{equation}
		\mathbf{r}_i^\mathrm{T}\mathbf{p}_{i}^\mathrm{c}>0,
	\end{equation}
	which indicates that the angle between $ \mathbf{r}_i $ and $ \mathbf{p}_{i}^\mathrm{c} $ is less than $\frac{\pi}{2}$, meaning the $ i $-th triangular mesh is facing towards the BS.
	\item{\textbf{Unobstructed Path Condition:}} This condition verifies that an unobstructed path exists between the BS and the target mesh $\mathcal{T}_i$. This is crucial because other objects could occlude the path, making a reflection impossible. To verify this, we perform a ray-triangle intersection test. We test rays from the BS to each mesh vertice for intersections using the M{\"o}ller--Trumbore algorithm \cite{ray_tracing}. This verification process is implemented by evaluating a function $\mathrm{Inters}(\mathbf{v}_i^k, \mathcal{T}_j)$, which takes the vertex and the mesh as inputs and returns 1 if an intersection occurs, or 0 otherwise\footnote{Computation details can be found in \cite{ray_tracing}.}. Thus, the unobstructed path condition can be expressed as
	\begin{equation}
		\sum_{j\neq i} \mathrm{Inters}(\mathbf{v}_i^k, \mathcal{T}_j)=0,~\exists k \in \{1, 2, 3\},
	\end{equation}
	which indicates that for each $ i $-th triangular mesh, if there exists a ray which is not blocked by any other mesh, such mesh is defined as a valid mesh for VBS computation.
\end{itemize}

\textcolor{black}{After verifying these two conditions, the location of VBS indexed by $ (1,i) $ is computed as the image of the BS's location across the triangular mesh
\begin{gather}
	\tilde{\mathbf{o}}_\mathrm{VBS}^{1,i} = \mathbf{o}_\mathrm{BS} - 2d_i \mathbf{r}_{i},	
\end{gather}
where $ d_i $ is the distance between the BS and the $ i $-th triangular mesh.} However, there could exist an abundant number of VBSs of the same order caused by deviation in the geometric information of LiDAR points. Thus, it is necessary to reduce the number of VBSs. The locations of the original VBSs that are spatially close can be clustered to obtain a sparse representation of the environment. 
Thus, HDBSCAN is adopted again for VBS clustering to handle different densities of the original VBSs. The location of the $v$-th clustered 1-order VBS is determined as the centroid of the cluster, denoted by $ \mathbf{o}_\mathrm{VBS}^{1,v} $. The triangular meshes associated with this VBS are obtained by taking the union of all meshes corresponding to the original VBSs within the cluster. Higher order VBSs' locations can be computed similarly.

\subsubsection{Coverage Computation and Storage}
\textcolor{black}{To compute and store the coverage area, the entire region is first divided into $D_\mathrm{x} \times D_\mathrm{y}$ grid points, where the size can be adjusted for different storage requirements. For the BS's LoS coverage area $\mathcal{C}_{0,1}$, we traverse all grid points and determine whether a LoS path exists between the BS and each grid point by checking if the connecting line segment is obstructed by any objects in the environment. For the VLoS coverage area of VBS indexed by $(1,v)$ where $v>1$, the computation process is as follows. We traverse each grid point within the region and compute the intersection point between the line segment connecting the grid point and the VBS location $\mathbf{o}_\mathrm{VBS}^{1,v}$ with respect to the triangular meshes associated with this VBS. 
Once the intersection point is obtained, we examine the line segment between this intersection point and the grid point to determine whether it is obstructed by any other triangular meshes using the same method in Section \ref{vbs_location_computation}. If no obstruction is detected, the grid point is within the VLoS coverage area $\mathcal{C}_{1,v}$ of VBS $(1,v)$.}
\begin{remark}
	The main differences between the proposed VBS principles and existing CKM methods are threefold. VBS stores only integer VBS indexes and per-grid location information, rather than the high-precision complex channel parameters used in CKM \cite{ckm_beam_alignment}. Moreover, VLoS coverage maps can be reconstructed from acquired LiDAR at arbitrary precision. Finally, if the BS's location changes, CKM needs costly re-measurement whereas VBSs can be recomputed from existing LiDAR plus the new BS position.
\end{remark}

\subsection{VBS-assisted Beam Alignment with Coarse Channel Reconstruction}
\textcolor{black}{In beam alignment, the goal is to identify the beam pair with the largest virtual channel gain. Under the specular assumption, the estimated pair can be suboptimal, requiring partial beam training for refinement. To reduce search overhead, different beam pairs are assigned varying priorities through a VBS-assisted coarse channel reconstruction method. The coarse estimation first computes the potential AoA, AoD, and distance from the VBS and UE locations, and then estimates the channel gain using an empirical path loss model. Notably, 3D LiDAR points are not stored during the two-stage estimation, as their features are mainly represented by the VBS set, ensuring low storage cost.}
\textcolor{black}{\subsubsection{Possible AoA, AoD and Distance Computation}
\label{aoa_aod_estimation}
Based on the UE's location and stored coverage map, the BS can identify the set of candidate VBSs that serve the UE. The BS can then proceed to compute the corresponding channel parameters. To facilitate this computation, we define the following position vectors. For a VBS indexed by $(1,v)$, let $ \mathbf{p}_\mathrm{BS}^{1,v}=\mathbf{o}_\mathrm{ref}^{1,v}-\mathbf{o}_\mathrm{BS}$ and $ \mathbf{p}_\mathrm{UE}^{1,v}=\mathbf{o}_\mathrm{ref}^{1,v}-\mathbf{o}_\mathrm{UE}$ denote the displacement vectors from the reflection point with respect to the VBS indexed by $ (1,v) $ to the BS and UE, respectively. The reflection point is computed as
\begin{gather}
	\begin{aligned}
	\mathbf{o}_\mathrm{ref}^{1,v} &= \mathbf{o}_\mathrm{VBS}^{1,v} + \frac{ \|\mathbf{o}_{\mathrm{BS}} - \mathbf{o}_{\mathrm{VBS}}^{1,v}\|_2^2 \left( \mathbf{o}_{\mathrm{UE}} - \mathbf{o}_{\mathrm{VBS}}^{1,v} \right) }{ 2 \left( \mathbf{o}_{\mathrm{BS}} - \mathbf{o}_{\mathrm{VBS}}^{1,v} \right)^{\mathrm{T}} \left( \mathbf{o}_{\mathrm{UE}} - \mathbf{o}_{\mathrm{VBS}}^{1,v} \right) }.
	\end{aligned}
\end{gather}
For LoS path, the vectors are similarly defined as $ \mathbf{p}_\mathrm{BS}^{0,1}=\mathbf{o}_\mathrm{UE}-\mathbf{o}_\mathrm{BS} $ and $ \mathbf{p}_\mathrm{UE}^{0,1}=\mathbf{o}_\mathrm{BS}-\mathbf{o}_\mathrm{UE} $. Let $ \Delta \phi^\mathrm{r}, \Delta \phi^\mathrm{t} $ denote the inherent antenna array azimuth angle offsets at the UE and BS, respectively, and $ \Delta \psi^\mathrm{r}, \Delta \psi^\mathrm{t} $ denote the inherent antenna array elevation angle offsets at the UE and BS, respectively. We can compute the channel parameters corresponding to the VBS indexed by $ (u,v) $, i.e., azimuth/elevation AoA, AoD $ \hat{\phi}_{u,v}^\mathrm{r}, \hat{\phi}_{u,v}^\mathrm{t}, \hat{\psi}_{u,v}^\mathrm{r}, \hat{\psi}_{u,v}^\mathrm{t} $, and the two segment distances $ \hat{r}_{u,v}^\mathrm{r}, \hat{r}_{u,v}^\mathrm{t} $ as follows.
\begin{itemize}
    \item \textbf{For VLoS paths:} 
    $\hat{\phi}_{1,v}^\mathrm{r} = \arctan\left(\frac{\{\mathbf{p}_\mathrm{UE}^{1,v}\}_2}{\{\mathbf{p}_\mathrm{UE}^{1,v}\}_1}\right) + \Delta \phi^\mathrm{r}$,
    $\hat{\phi}_{1,v}^\mathrm{t} = \arctan\left(\frac{\{\mathbf{p}_\mathrm{BS}^{1,v}\}_2}{\{\mathbf{p}_\mathrm{BS}^{1,v}\}_1}\right) + \Delta \phi^\mathrm{t}$,
    $\hat{\psi}_{1,v}^\mathrm{r} = \arccos\left(\frac{\{\mathbf{p}_\mathrm{UE}^{1,v}\}_3}{\|\mathbf{p}_\mathrm{UE}^{1,v}\|_2}\right) + \Delta \psi^\mathrm{r}$,
    $\hat{\psi}_{1,v}^\mathrm{t} = \arccos\left(\frac{\{\mathbf{p}_\mathrm{BS}^{1,v}\}_3}{\|\mathbf{p}_\mathrm{BS}^{1,v}\|_2}\right) + \Delta \psi^\mathrm{t}$,
    $\hat{r}_{1,v}^\mathrm{r} = \|\mathbf{p}_\mathrm{UE}^{1,v}\|_2$, $\hat{r}_{1,v}^\mathrm{t} = \|\mathbf{p}_\mathrm{BS}^{1,v}\|_2$.
    \item \textbf{For LoS path:} 
    $\hat{\phi}_{0,1}^\mathrm{r} = \arctan\left(\frac{\{\mathbf{p}_\mathrm{UE}^{0,1}\}_2}{\{\mathbf{p}_\mathrm{UE}^{0,1}\}_1}\right) + \Delta \phi^\mathrm{r}$,
    $\hat{\phi}_{0,1}^\mathrm{t} = \arctan\left(\frac{\{\mathbf{p}_\mathrm{BS}^{0,1}\}_2}{\{\mathbf{p}_\mathrm{BS}^{0,1}\}_1}\right) + \Delta \phi^\mathrm{t}$,
    $\hat{\psi}_{0,1}^\mathrm{r} = \arccos\left(\frac{\{\mathbf{p}_\mathrm{UE}^{0,1}\}_3}{\|\mathbf{p}_\mathrm{UE}^{0,1}\|_2}\right) + \Delta \psi^\mathrm{r}$,
    $\hat{\psi}_{0,1}^\mathrm{t} = \arccos\left(\frac{\{\mathbf{p}_\mathrm{BS}^{0,1}\}_3}{\|\mathbf{p}_\mathrm{BS}^{0,1}\|_2}\right) + \Delta \psi^\mathrm{t}$,
    $\hat{r}_{0,1}^\mathrm{r} = \hat{r}_{0,1}^\mathrm{t} = \|\mathbf{p}_\mathrm{BS}^{0,1}\|_2$.
\end{itemize}
}

\subsubsection{Path Loss Computation}
\textcolor{black}{The path loss estimation can be divided into two parts, the free space path loss and reflection path loss. Diffraction loss is neglected since it has hardly no contribution to the stable wireless communication. The free space path loss can be computed as} 
\begin{gather}
	\mathrm{PL}_\mathrm{free}(f_\mathrm{c}, d) = 20\log_{10}\left(\frac{4\pi f_\mathrm{c}}{c}\right) + 20\log_{10}(d),
	\label{free_space_pathoss}
\end{gather}
where $ f_\mathrm{c} $ is the carrier frequency, $ c $ is the speed of light, $ \lambda $ is the wavelength, and $ d $ is the path distance. \textcolor{black}{The $ u $ times reflection path loss can be estimated as}
\begin{gather}
	\mathrm{PL}_\mathrm{ref}(f_\mathrm{c}, d, u, \Gamma) = \mathrm{PL}_\mathrm{free}(f_\mathrm{c}, d) + \Gamma [\text{dB}] + 10\log_{10}u,
	\label{reflection_pathoss}
\end{gather}
where $ \Gamma $ is a predefined reflection coefficient since the material properties are unknown.



\textcolor{black}{For VBS indexed by $ (u,v) $, there exists a path with $ u $ times reflections, the channel gain can be estimated as $ \hat{\beta}_{u,v} = 10 ^ {\left(\frac{-\mathrm{PL}_{u,v}}{20}\right)} $, where $ \mathrm{PL}_{u,v} $ is computed as (\ref{total_path_loss}), shown at the top of the next page.} 
\begin{figure*}
	\centering
	\begin{gather}
		\mathrm{PL}_{u,v}=\mathbbm{1}(u=0)\mathrm{PL}_\mathrm{free}(f_\mathrm{c}, \|\mathbf{p}_\mathrm{BS}^{0,1}\|_2)+(1-\mathbbm{1}(u=0))\mathrm{PL}_\mathrm{ref}(f_\mathrm{c}, \|\mathbf{p}_\mathrm{UE}^{1,v}\|_2+\|\mathbf{p}_\mathrm{BS}^{1,v}\|_2, 1, \Gamma),
		\label{total_path_loss}
	\end{gather}
	\hrule 
\end{figure*}
With the azimuth/elevation AoA, AoD, and segment distances, the coarse channel between BS and UE can be reconstructed based on the geometric model as
\begin{small}
	\begin{gather}
		\begin{aligned}
			\hat{\mathbf{H}} = &\sqrt{N_\mathrm{BS}N_\mathrm{UE}}\sum_{(u,v)\in \mathcal{V}_\mathrm{UE}}\hat{\beta}_{u,v} \times\\
			&\mathbf{a}(\hat{\psi}_{u,v}^\mathrm{r}, \hat{\phi}_{u,v}^\mathrm{r}, \hat{r}_{u,v}^\mathrm{r}; N_\mathrm{UE})\mathbf{a}(\hat{\psi}_{u,v}^\mathrm{t}, \hat{\phi}_{u,v}^\mathrm{t}, \hat{r}_{u,v}^\mathrm{t}; N_\mathrm{BS})^\mathrm{H},
		\end{aligned}
		\label{reconstructed_channel}
	\end{gather}
\end{small}
where $\mathcal{V}_\mathrm{UE}$ denotes the VBS set serving the UE, which can be obtained by first mapping the UE's location to the $K$ nearest grid points $(i_\mathrm{x}, i_\mathrm{y})$, and then taking the union of the VBS sets corresponding to each of these $K$ grid points. $ \mathbf{a}(\hat{\psi}_{u,v}^\mathrm{r}, \hat{\phi}_{u,v}^\mathrm{r}, \hat{r}_{u,v}^\mathrm{r}; N_\mathrm{UE}) $ and $ \mathbf{a}(\hat{\psi}_{u,v}^\mathrm{t}, \hat{\phi}_{u,v}^\mathrm{t}, \hat{r}_{u,v}^\mathrm{t}; N_\mathrm{BS}) $ are UE-side and BS-side steering vectors, respectively. 
The steering vector of near-field ULA with $ N $ elements and carrier frequency $ f_\mathrm{c} $ is
\begin{small}
	\begin{gather}
		\mathbf{a}(\psi, \phi, r; N) = \frac{1}{\sqrt{N}} \left[ e^{jk_\mathrm{c} (r^{(0)} - r)}, \ldots, e^{jk_\mathrm{c} (r^{(N-1)} - r)} \right]^\mathrm{T},
	\end{gather}
\end{small}
where $ k_\mathrm{c}=\frac{2\pi f_\mathrm{c}}{c} $ is wave number, $ c $ is speed of light, and $ r^{(n)} $ is distance between $ n $-th antenna and scatter, defined as $ r^{(n)} = \sqrt{r^2 + (\delta_n d)^2 - 2 r \sin\psi \sin\phi \delta_n d} $. $ \delta_n = n - \frac{N-1}{2} $ is array-centered normalized antenna index, $ d $ is antenna spacing, $ r $ is distance between scatter and array. 

Predefined codebook serves as transformation matrix to get polar domain beamspace representation $ \mathbf{G} = |\mathbf{U}_\mathrm{UE}^\mathrm{H}\hat{\mathbf{H}}\mathbf{U}_\mathrm{BS}| $. Top-$ S $ beam pair indexes with the largest virtual channel gains form searching subset $ \mathcal{S} $. Partial beam training based on $ \mathcal{S} $ yields optimal beam pair $ (b_\mathrm{BS}^\star, b_\mathrm{UE}^\star) $ with maximum virtual channel gain.

\section{Results and Discussion}
\label{evaluation}


\subsection{Simulation Setup}
\begin{table}[t!]
	\footnotesize
	\renewcommand{\arraystretch}{1}
	\centering
	\caption{Scenario-dependent dataset configuration.}
	\label{setup_config}
	\begin{tabular}{cc}
		\toprule
		\textbf{Parameter} & \textbf{Value} \\
		\midrule
		Carrier Frequency $ f_\mathrm{c} $ & 40 GHz \\
		\hline
		Bandwidth $ W $ & 500 MHz \\
		\hline
		BS Position $ \mathbf{o}_\mathrm{BS} $ & [140, 60, 4] m \\
		\hline
		UE's height $ h_\mathrm{UE} $ & 1.5 m \\
		\hline
		Transmit Power $ P_\mathrm{T} $ & 40 dBm \\
		\hline
		Noise Power Spectrum Density $ N_0 $ & -174 dBm/Hz \\
		\hline
		Predefined Reflection Coefficient $ \Gamma $ & 10 \\
		\hline
		$ K $ for Grid Neighborhood & 3\\
		\hline
		VBS Grid Number $ D_\mathrm{x}, D_\mathrm{y} $ & 40, 40\\
		\bottomrule
	\end{tabular}
\end{table}
In this section, we evaluate the SE performance of beam alignment with the proposed VBS technique. The simulation is conducted in a 3D environment with the help of Blender  and Sionna \cite{sionna}. Blender is used to import the OSM in a specific region in Mexia, Texas with about 120 m $ \times $ 120 m ($ 31.6802^\circ\text{-}31.6823^\circ\text{N}$, $ -96.4826^\circ\text{-}-96.4787^\circ\text{W} $), and generate simulation 3D LiDAR points, while Sionna is used to simulate the ray-tracing based wireless channels. To simulate the LiDAR measurement and process error, the standard deviation and drop rate are set to 5 cm and 5\%. We consider a sub-THz fully-analog massive MIMO system with configuration in Table \ref{setup_config}. 


\subsection{Results and Discussion}
\begin{figure}[t!]
	\centering
	\includegraphics[width=0.78\linewidth]{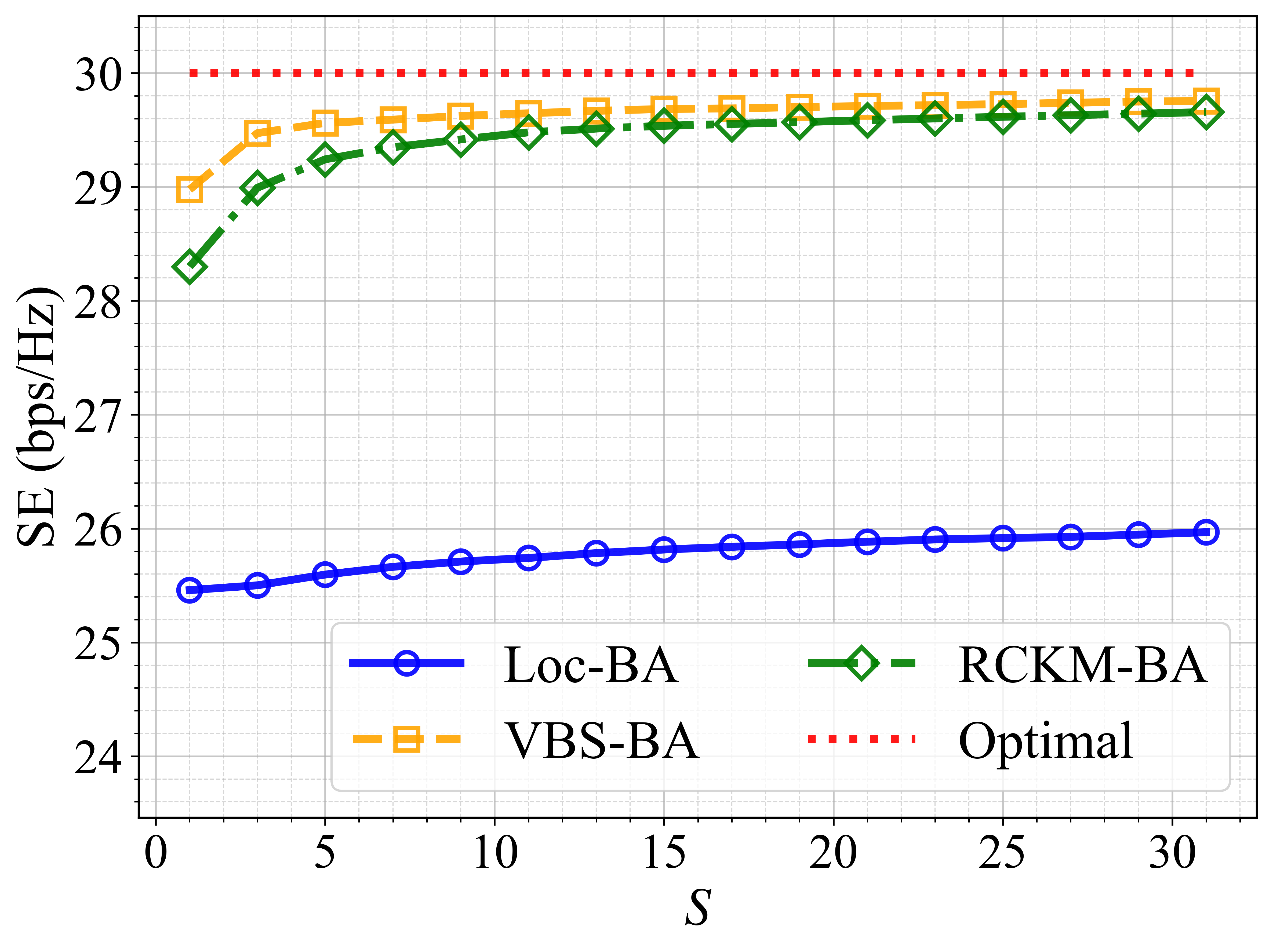}
	\caption{SE performance versus Top-$ S $ partial beam training number under the $  N_\mathrm{BS}=256, N_\mathrm{UE}=16 $ configuration.}
	\label{su_mimo_se_beam_training}
\end{figure}
We compare the performance of VBS-assisted beam alignment (VBS-BA) with location based beam alignment (Loc-BA) \cite{ckm_beam_alignment} and refined CKM-based beam alignment (RCKM-BA). When $ S = 1 $, the RCKM-BA degrades into a method called training-free beam alignment in \cite{ckm_beam_alignment}. The grid number for storing the first 4 strongest channel paths parameters is $ 40 \times 40 $. \textcolor{black}{The optimal line means choosing the best beam pair to maximize the virtual channel gain by exhaustive beam training.}

As shown in Fig. \ref{su_mimo_se_beam_training}, under the $ N_\mathrm{BS}=256, N_\mathrm{UE}=16 $ configurations, the Loc-BA performs much worse than VBS-BA and RCKM-BA, because it is only based on the LoS path. The coarse reconstructed channel only based on location data cannot capture dominant NLoS paths. VBS-BA has a little performance degradation compared to the optimum. Such acceptable gap is caused by neglect of coherence stack, errors in coverage area computation and measurement errors in 3D LiDAR data. If partial beam training number is less than 5, VBS-BA can surpass up to 0.5 bps/Hz compared to RCKM-BA due to the limited resolution of CKM measurement. 
\begin{table}[t!]
\footnotesize
\renewcommand{\arraystretch}{1.}
\centering
\caption{SE performance versus different antenna configuration with Top-5 partial beam training.}
\label{su_mimo_se_antenna_configuration}
\begin{tabular}{ccccc}
\toprule
\multirow{2}{*}{\makecell{\textbf{Antennas for}\\\textbf{BS and UE}}} & \multicolumn{4}{c}{\textbf{SE (bps/Hz)}} \\
\cline{2-5}
& \textbf{Loc-BA} & \textbf{VBS-BA} & \textbf{RCKM-BA} & \textbf{Optimal} \\
\midrule
(256, 16) & 25.60 & 29.56 & 29.24 & {30.00} \\
(256, 32) & 25.51 & 30.00 & 29.69 & {30.48} \\
(512, 8) & 25.57 & 29.08 & 28.69 & {29.85} \\
(512, 16) & 25.58 & 29.63 & 29.15 & {30.36} \\
\bottomrule
\end{tabular}
\end{table}
From Table \ref{su_mimo_se_antenna_configuration}, with different antenna configuration, VBS-BA always achieves the best performance compared to Loc-BA and RCKM-BA with Top-5 partial beam training. To get the optimal solution, the exhaustive beam training overhead should reach $  896 \times 20,  896 \times 51, 2412 \times 8, 2412 \times 20 $ based on the near-field beam codebook in \cite{channel_estimation_far_near_field}, respectively. As shown in Fig. \ref{su_mimo_se_beam_training} and Table \ref{su_mimo_se_antenna_configuration}, for VBS-BA, only training Top-5 beam pairs achieves about 98\% SE with little performance loss, and training more beam pairs only brings incremental increase in SE.

\section{Conclusion}
\label{conclusion}
In this paper, we proposed a  VBS construction technique to represent the main features of the wireless environment with regional 3D LiDAR points and location data. The VBSs can be used to assist partial beam training by coarse channel reconstruction with arbitrary resolution. The estimated absolute value of the channel gain is based on LoS and 1-order specular reflection under empirical models, denoting the importance of each path for subsequent effective partial beam training. Simulation results based on a real-world map and ray-tracing based channel demonstrated that VBS-assisted beam alignment achieves about 98\% performance of the optimal result by training only Top-5 beams.

\normalem
\bibliographystyle{IEEEtran}
\bibliography{IEEEabrv,reference}

\newpage
\clearpage

\end{document}